\documentclass{iau}
\usepackage{amsmath}
\usepackage{natbib}
\usepackage{graphicx}
\usepackage{wrapfig}
\usepackage{caption}
\usepackage{subcaption}
\usepackage{booktabs}
\usepackage{verbatim}
\usepackage{lineno}
\usepackage{adjustbox}
\usepackage{amssymb}
\usepackage[utf8]{inputenc}
\usepackage[english]{babel}

\begin{document}

\lefttitle{P. Moraga Baez et al.}
\righttitle{ALMA Observations of Bipolar PNe}

\journaltitle{Planetary Nebulae: a Universal Toolbox in the Era of Precision Astrophysics}
\jnlDoiYr{2023}
\doival{10.1017/xxxxx}
\volno{384}

\aopheadtitle{Proceedings IAU Symposium}
\editors{O. De Marco, A. Zijlstra, R. Szczerba, eds.}

\title{ALMA Observations of Molecular Line Emission from High-excitation Bipolar Planetary Nebulae}

\author{Paula Moraga Baez$^1$, Joel H. Kastner$^{1,2}$, Jesse Bublitz$^3$, Javier Alcolea$^4$, Miguel Santander-Garcia$^4$, Thierry Forveille$^5$, Pierre Hily-Blant$^5$, Bruce Balick$^6$, Rodolfo Montez, Jr.$^7$, Caroline Gieser$^8$}
\affiliation{$^1$School of Physics and Astronomy and Laboratory for Multiwavelength Astrophysics, Rochester Institute of Technology, Rochester, NY, USA}
\affiliation{$^2$Chester F. Carlson Center for Imaging Science, Rochester Institute of Technology}
\affiliation{$^3$Green Bank Observatory, Green Bank, WV, USA}
\affiliation{$^4$Observatorio Astronómico Nacional, Madrid, Spain}
\affiliation{$^5$Institut de Planetologie et d'Astrophysique de Grenoble, France}
\affiliation{$^6$Department of Astronomy, University of Washington, Seattle, WA USA}
\affiliation{$^7$Center for Astrophysics, Harvard \& Smithsonian, Cambridge, MA, USA}
\affiliation{$^8$Max Planck Institute for Extraterrestrial Physics, Munich, Germany}

\begin{abstract}
We present early results from our program of ALMA Band 6 (1.3mm) molecular line mapping of a sample of nearby, well-studied examples of high-excitation, bipolar/pinched-waist and molecule-rich planetary nebulae (Hubble 5 and NGC 2440, 2818, 2899, 6302, and 6445). 
We have mapped these planetary nebulae (PNe) in isotopologues of CO as well as various molecular line tracers of high-energy irradiation, such as HCN, CN, HNC, and HCO$^+$, 
with the complementary goals of establishing nebular kinematics as well as the zones of UV-heated and X-ray-ionized molecular gas within each nebula. The resulting high-resolution ALMA molecular emission-line maps reveal the regions of high-excitation bipolar PNe in which molecular gas, presumably ejected during asymptotic giant branch stages of the PN progenitor stars, survives and evolves chemically. We present a summary of molecular species detected to date in the sample nebulae, and we use example results for one PN (NGC 6455) to demonstrate the power of the ALMA data in revealing the structures, kinematics, and compositions of the equatorial molecular tori that are a common feature of the sample objects.
\end{abstract}


\maketitle

\vspace{-4mm}
\section{Introduction} \label{sec:intro}

For several decades, mm-wave CO emission lines have been used as probes of the molecular gas components of planetary nebulae \citep[PNe; e.g.,][]{Huggins1989}. Numerous single-dish mm-wave radio telescope surveys targeting CO, HCO$^{+}$, HNC, HCN, CN, CS, and other species have been carried out so as to advance our understanding of the molecular masses and chemistries of PNe \citep[e.g.,][]{Huggins2000,Edwards2013,Edwards2014,Bublitz2019,Schmidt2022}. Interferometric mm-wave molecular line mapping observations of PNe are somewhat fewer and farther between, and have usually involved mapping of individual PNe \citep[e.g.,][]{Bublitz2023}.

Here, we present initial results from Atacama Large Millimeter Array (ALMA) Band 6 (1.3 mm) observations of a sample of relatively nearby, molecule-rich, high-excitation bipolar PNe. This program constitutes among the first molecular line surveys of PNe to exploit ALMA's unique high spatial and spectral resolution mapping capabilities. The main goals of our ALMA bipolar PN survey are twofold: {\it (1)} to use velocity-resolved mm-wave molecular line mapping to pinpoint the locations and study the structures and kinematics of the regions of cold, dense molecular gas within high-excitation bipolar PNe; and {\it (2)} to ascertain the effects of UV vs. X-ray irradiation from PN central stars on the composition and heating of PN molecular gas. 

Ultimately, these observations of an evolutionary sequence of bipolar PNe ranging from young and rapidly evolving nebulae to more ``mature" bi-lobed nebulae will help us constrain PN shaping models that involve binary-influenced AGB and post-AGB mass loss.

\section{Observations}

\subsection{Sample}

Our target nebulae (Table~\ref{tbl:PNprops}) have been selected from comprehensive catalogs of solar neighborhood PNe \citep{Cahn1992,Frew2013,Stanghellini2002} on the basis of ({\it i}) distances $D\lesssim$2 kpc; ({\it ii}) pinched-waist bipolar morphologies; ({\it iii}) large molecular gas masses, as evidenced by bright near-IR H$_2$ and/or mm-wave CO emission \citep{Fang2018,Guerrero2000,Huggins1996,Huggins2005,Kastner1996,Phillips1992}; and ({\it iv}) hot, luminous central stars (i.e., central star effective temperatures $T_{\rm eff} \gtrsim 150$ kK and luminosities $L_\star \gtrsim 10^3 L_\odot$). All of the Table~\ref{tbl:PNprops} objects are hence likely the descendants of relatively massive progenitor stars, with initial masses in the range  $\sim$3--8 $M_\odot$ \citep{CorradiSchwarz1995,Kastner1996,KarakasLugaro2016,MillerBertolami2016}. 



\begin{table}
\begin{center}
\caption{\sc Properties of Survey Targets}
\vspace{.1in}
\label{tbl:PNprops}
\footnotesize
\begin{tabular}{cc|cc|ccc}
\toprule
& & \multicolumn{2}{c}{\sc Central Star$^a$} & \multicolumn{3}{c}{\sc Molecular Torus$^b$}\\
Object & $D^a$ & $\log L_{\star}$ & $T_{\rm eff}$ & $R$ & $V_{\rm exp}$ & Age\\
& (kpc) & ($L_\odot$) & (kK) & (pc) & (km s$^{-1}$) & (yr)\\
\midrule
\midrule
Hubble 5 & 1.7 & 3.64 & 170 & 0.033 & 50 & 1300\\
NGC 2440 & 1.9 & 3.32 & 210 & 0.05 & 35 & 2800\\
NGC 6302 & 1.0 & 3.62 & 220 & 0.05 & 20 & 5000\\
NGC 6445 & 1.14$^c$ & 2.80 & 170 & 0.11 & 55 & 5000\\
NGC 6537 & 2.0 & 3.30 & 180 & 0.08 & 20 & 8000\\
NGC 2818 & 2.1 & 2.80 & 145 & 0.14 & 20 & 15000\\
NGC 2899 & 1.9$^c$ & 4.25$^d$ & 270$^d$ & 0.16 & 20 & 17000\\
\end{tabular}
\end{center}
\footnotesize
{\sc Notes:} 
a) Distances ($D$) and central star properties ($\log L_\star$, $T_{\rm eff}$) as listed in \citet{Frew2013} except where noted. b) Radius ($R$), deprojected expansion velocity ($V_{\rm exp}$), and dynamical age of molecular torus as estimated from ALMA $^{12}$CO (2--1) data. c) $D$ from Gaia DR3. d) $\log L_{\star}$ and $T_{\rm eff}$  from \citet{GonzalezSantamaria2019}.
\end{table}
\vspace{-5mm}
\subsection{ALMA data}

Our ALMA Band 6 survey, conducted under ALMA programs 2021.1.00456.S, 2021.2.00004.S, and 2022.1.00401.S, used a half-dozen spectral setups in the 220--270 GHz frequency range to target emission lines of CO, CN, HCN, HNC, HCO$^{+}$, CO$^{+}$, CS, SO, and isotopologues of CO, HCN, and HCO$^{+}$ in the sample PNe. By observing molecular emission from these molecular species, we seek to determine the dynamical ages (ejection timescales) of their molecular emitting regions and (hence) AGB-terminating mass loss, and we can place constraints on the molecular gas optical depths, PN progenitor masses, and zones of UV-irradiated and X-ray-ionized molecular gas \citep[see, e.g.,][and references therein]{Bublitz2019,Bublitz2023}. 
Table~\ref{tbl:allObs} lists the molecular lines targeted in the survey. We used both ALMA's 12-m array and Atacama Compact Array (ACA) in configurations that yielded $\sim$0.5$''$ and $\sim$5$''$ resolution imaging, respectively. Details of the ALMA array and spectrometer configurations and data processing, as well as results from archival ALMA Band 3 mapping of NGC 6537 \citep[obtained via ALMA program 2018.1.00424.S;][]{Gieser2023}, will be presented in forthcoming papers (Moraga Baez et al.\ 2024a,b, in prep.). Here, we present selected initial results from ALMA Band 6 observations of six Table~\ref{tbl:PNprops} PNe, as obtained from the ALMA ADMIT pipeline\footnote{https://admit.astro.umd.edu}: Hb 5 and NGC 2440, 2818, 2899, 6302, and 6445. 

\begin{table}
\begin{center}
\caption{\sc Band 6 Targeted Molecular Lines}
\vspace{.1in}
\label{tbl:allObs}
\footnotesize
\begin{adjustbox}{width=1\textwidth}
\small
\begin{tabular}{lccccccc}
\toprule
Molecule (Trans.) & $\nu$ (GHz) & Hb 5 & NGC 6302 & NGC 2440 & NGC 6445 & NGC 2818 & NGC 2899\\
\midrule
\midrule
$^{12}$CO (2--1) & 230.538 & $\checkmark$ & $\checkmark$ & $\checkmark$ & $\checkmark$ & $\checkmark$ & $\checkmark$\\
$^{13}$CO (2--1) & 220.398 & O & $\checkmark$ & $\checkmark$ & O & $\checkmark$ & $\checkmark$\\
C$^{18}$O (2--1) & 219.560 & O & $\checkmark$ & $\times$ & O & $\times$ & $\times$\\
CS (5--4) & 244.936 & $\times$ & $\checkmark$ & $\times$ & $\checkmark$ & $\checkmark$ & $\times$\\
C$^{34}$S (5--4) & 241.016 & O & $\checkmark$ & $\times$ & O & $\times$ & $\times$\\
CN (2--1) & 226.697$^b$,226.874$^b$ & $\checkmark$ & $\checkmark$ & $\checkmark$ & $\checkmark$ & $\checkmark$ & $\checkmark$\\
HCN (3--2) & 265.886 & $\checkmark$ & $\checkmark$ & $\checkmark$ & O & $\checkmark$ & $\checkmark$\\
H$^{13}$CN (3--2) & 259.011 & O & $\checkmark$ & $\checkmark$ & $\times$ & $\times$ & $\times$\\
HNC (3--2) & 271.981 & O & $\checkmark$ & $\checkmark$ & $\checkmark$ & $\checkmark$ & $\checkmark$\\
HCO$^+$ (2--1) & 267.557 & $\checkmark$ & $\checkmark$ & $\checkmark$ & O & $\checkmark$ & $\checkmark$\\
H$^{13}$CO$^+$ (3--2) & 260.255 & $\checkmark$ & $\checkmark$ & $\checkmark$ & $\times$ & $\times$ & $\times$\\
CO$^+$ (2--1) & 235.789$^b$, 236.063$^b$ & O & $\times$ & $\times$ & O & $\times$ & $\times$\\
SO (5$_6$--4$_5$) & 251.857 & O & $\checkmark$ & $\times$ & O & $\times$ & $\times$\\
SO (6$_5$--5$_4$) & 219.949 & O & $\checkmark$ & $\checkmark$ & O & $\times$ & $\times$\\
\end{tabular}
\end{adjustbox}
\end{center}
\footnotesize
{\sc Notes:} 
a) $\checkmark=$ detected; $\times=$ not detected; O $=$ data not obtained. b) Hyperfine complex; representative
transition frequency listed.
\end{table}

\section{Initial Survey Results: Overview} \label{sec:data}

\begin{figure}[b!]
    \centering
    \includegraphics[width=0.99\textwidth]{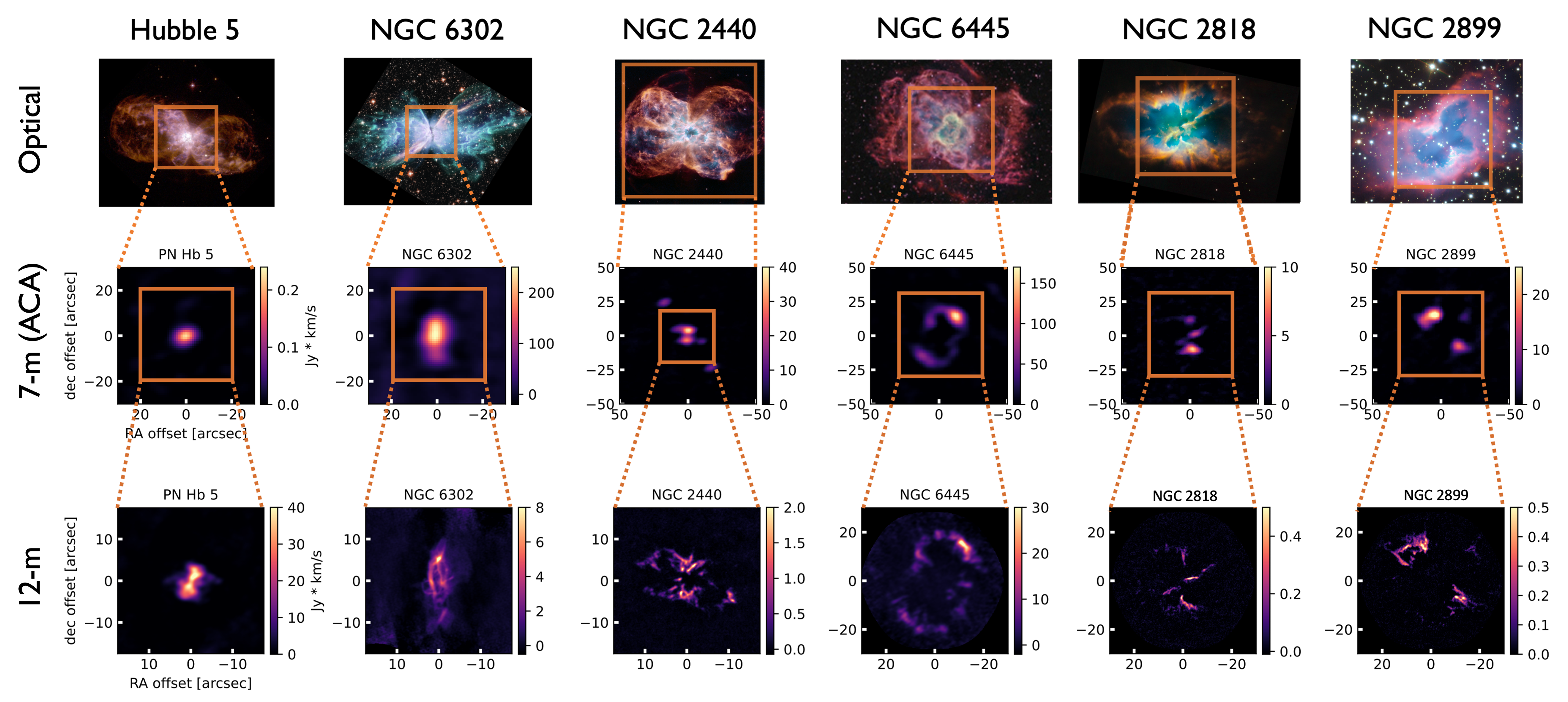}
    \caption{Comparison of optical and ALMA $^{12}$CO images of (from L to R) PNe Hb 5, NGC 6302, NGC 2440, NGC 6445, NGC 2818, and NGC 2899. {\it Top row}: color composites of optical images of the six PNe. Archival HST images are presented for Hb 5, NGC 6302, NGC 2440, and NGC 6445; archival ESO and NOT images are presented for NGC 2899 and NGC 6445, respectively. {\it Middle} and {\it bottom rows}: Band 6 $^{12}$CO (2--1) velocity-integrated images obtained with ALMA's ACA and 12-m arrays, respectively.} 
    \label{fig:Band6PNe}
\end{figure}

Fig.~\ref{fig:Band6PNe} presents velocity-integrated ACA and 12-m array $^{12}$CO (2--1) maps for the six PNe. Archival H$\alpha$ images are displayed above the velocity-integrated $^{12}$CO (2--1) maps for comparison. It is immediately evident that the brightest $^{12}$CO (2--1) in each nebula is confined to its central, pinched-waist equatorial region. In the case of Hb 5, NGC 6302, NGC 2440, and NGC 6445, the $^{12}$CO emission appears to take the form of a molecular torus. In NGC 2818 and NGC 2899, however, the $^{12}$CO emission lies in more isolated structures (clumps) that appear to surround, or perhaps lie embedded within, the central, equatorial, UV-ionized zones of these nebulae. 

The $^{12}$CO (2-1) maps and spatially integrated spectral line profiles were used to estimate molecular torus dynamical ages for the sample PNe (Table~\ref{tbl:PNprops}, last column). It is noteworthy that the nebulae with the brightest and most spatially coherent molecular tori (Hb 5, NGC 6302, NGC 2440, and NGC 6445) have the youngest dynamical ages ($>$5000 yr); while  NGC 2818 and NGC 2899, the oldest PNe in our sample (dynamical ages of $\sim$15000 yr and $\sim$17000 yr, respectively), display faint, clumpy $^{12}$CO (2--1) emission. In the youngest nebulae (Hb 5, NGC 2440, NGC 6302), some $^{12}$CO (2--1) emission structures are observed to sit within the large ionized lobes, suggesting that in young bipolar PNe, some molecular gas survives in these more heavily irradiated, fast-expanding regions.

\begin{figure}
    \centering
    \includegraphics[width=\textwidth]{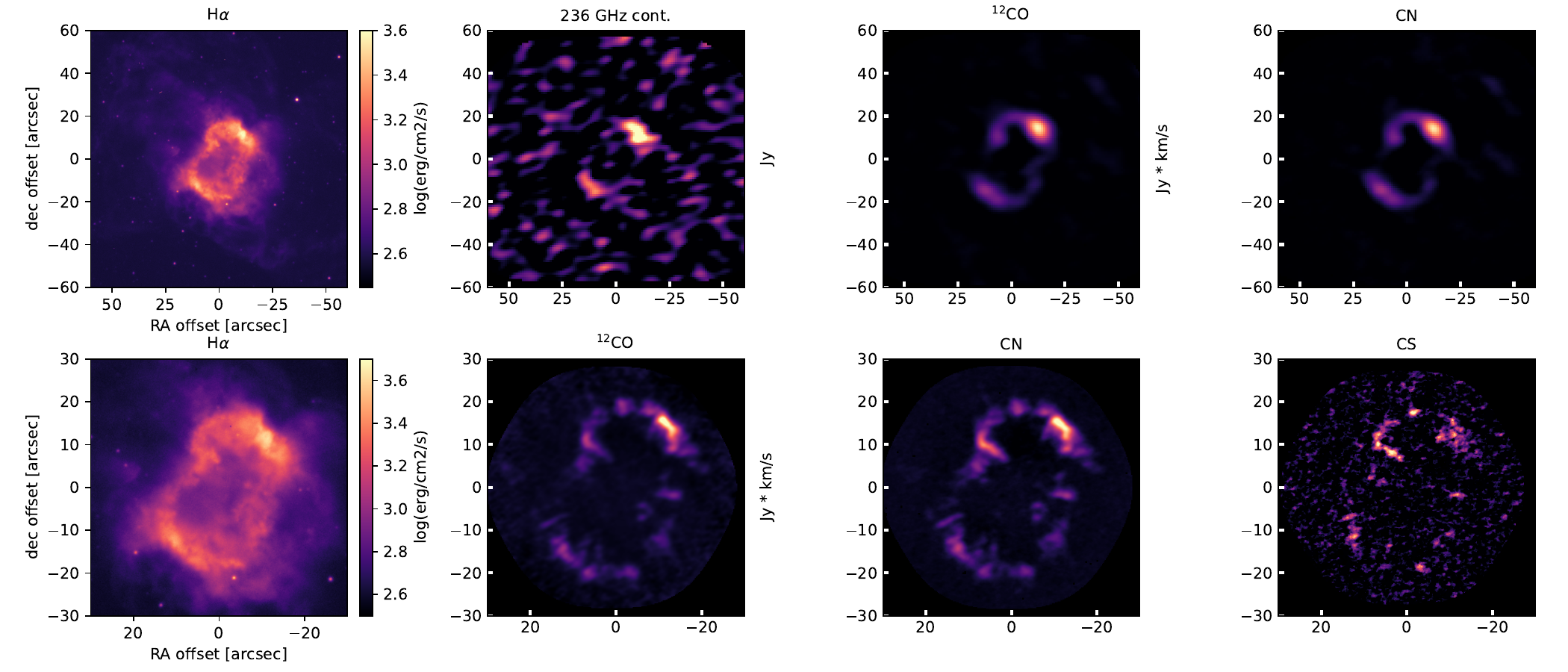}
    \caption{Below: Montages of ALMA continuum and velocity-integrated $^{12}$CO, CN and CS images of NGC 6445 as obtained with ACA ({\it top panels}) and the 12-m array ({\it bottom panels}). Each ALMA montage is presented alongside the NOT H$\alpha$ image \citep{Fang2018}, for reference. \vspace{2mm}}
    \label{fig:mom0Mont}
\end{figure}

\begin{figure}[ht]
    \centering
    \includegraphics[width=\textwidth]{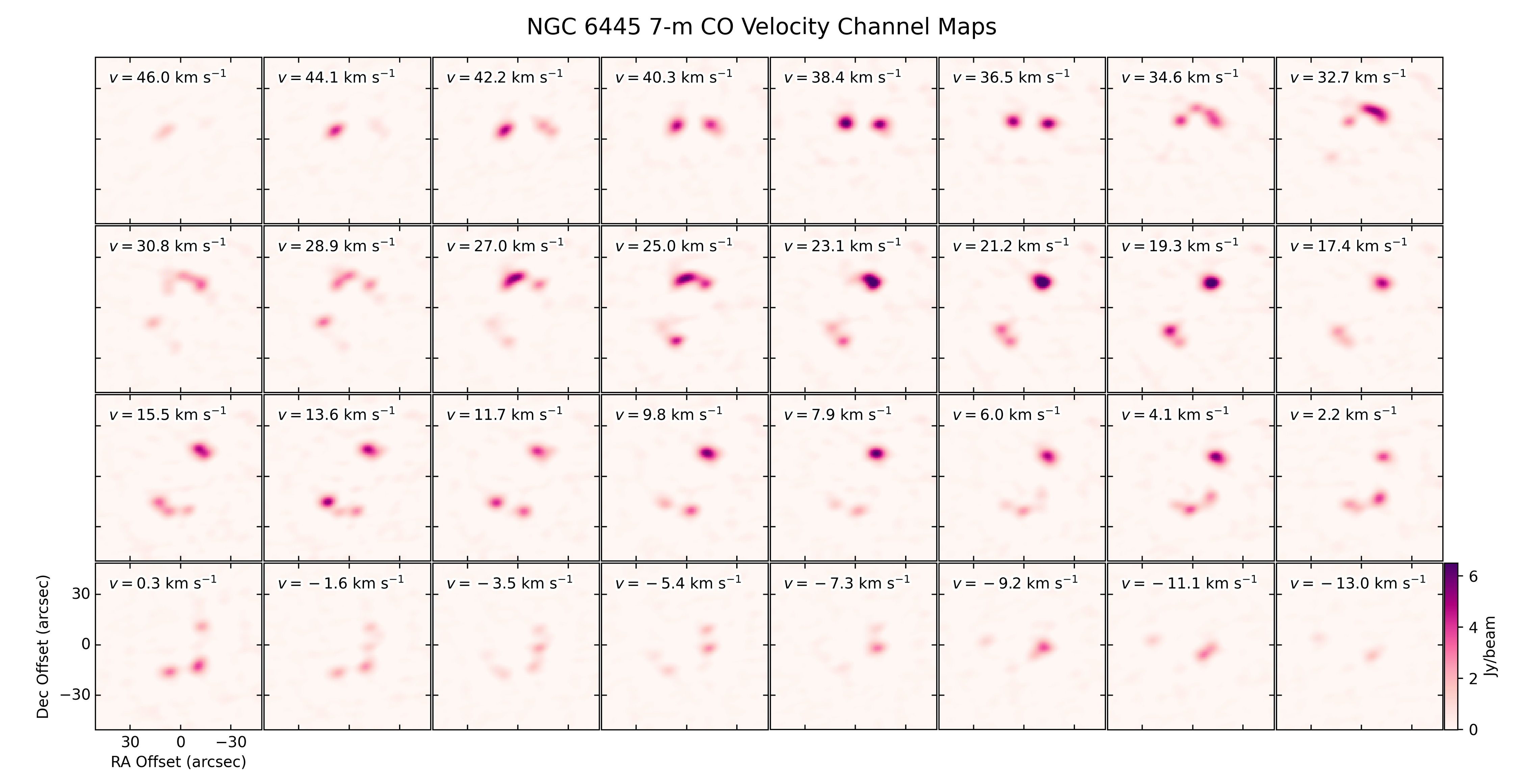}
    \caption{Velocity channel maps extracted from the NGC 6445 $^{12}$CO (2--1) ACA data cube. From left to right and top to bottom, the maps are displayed from $\sim$46 km s$^{-1}$ to $\sim-$13 km s$^{-1}$ in increments of 1.9 km s$^{-1}$.}
    \label{fig:VelocityChan}
\end{figure}

Table~\ref{tbl:allObs} summarizes the molecular emission lines that have been detected and not detected to date for each of the six PNe that were the subjects of ALMA Band 6 observations. The PNe are listed (from left to right) in order of increasing estimated dynamical age, and there appears to be a general trend wherein the detection of rare isotopologues (e.g., H$^{13}$CN, H$^{13}$CO$^+$) as well as S-bearing species (SO, CS) falls off from the youngest (NGC 6302) to oldest (NGC 2818, NGC 2899) objects. The lack of detections of rare isotopologues and S-bearing molecules in the older PNe likely reflects their relatively small residual molecular gas masses, although it remains to establish whether this potential trend might also be indicative of PN progenitor mass. 

\section{A Closer Look at NGC 6445}


\begin{wrapfigure}[28]{r}{0.5\textwidth}
  \centering
  \vspace{-10pt}
  \includegraphics[width=0.48\textwidth]{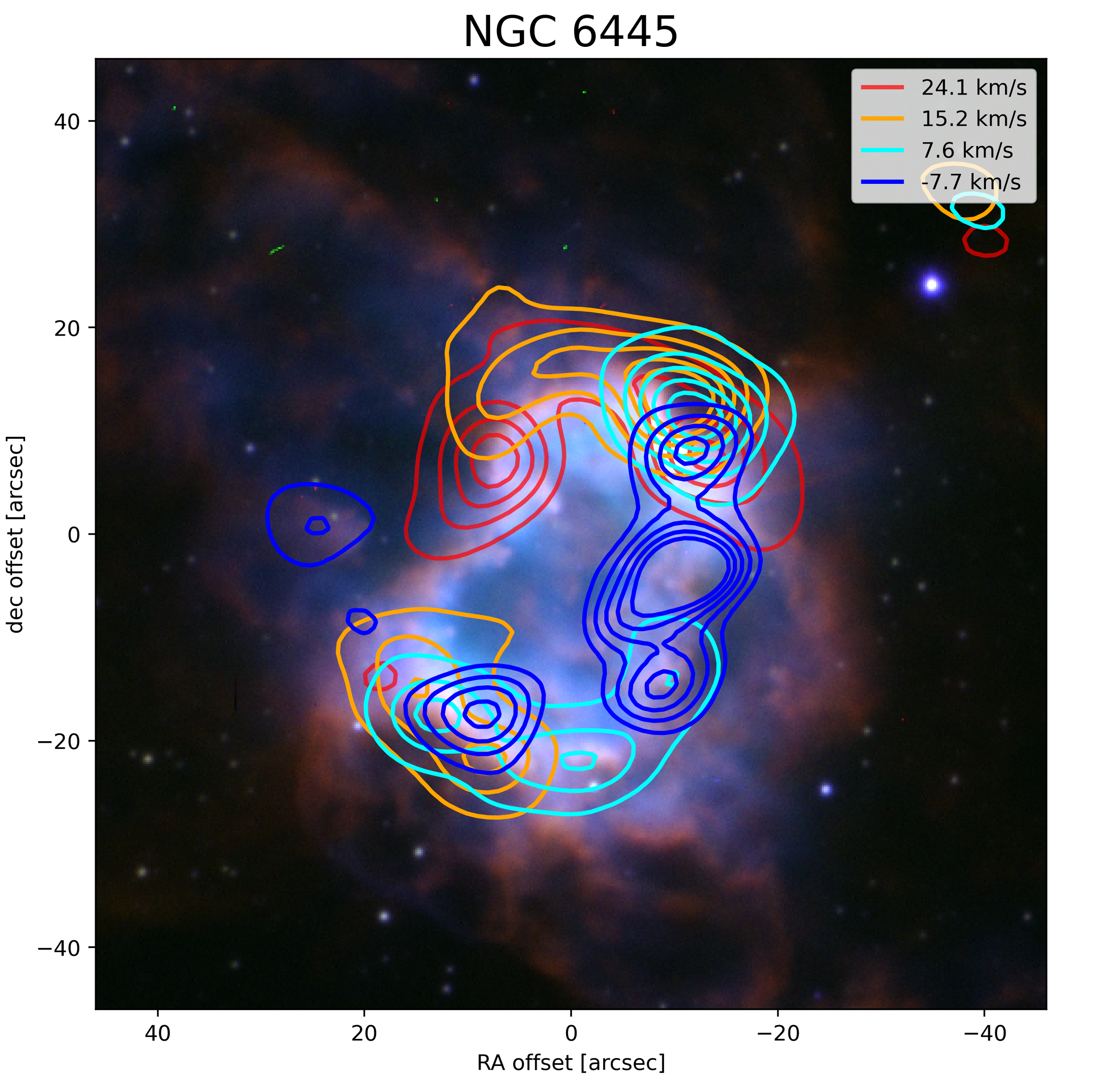}
  \caption{Contours of velocity-resolved $^{12}$CO (2--1) emission extracted from the NGC 6445 ACA data cube, overlaid on a color composite of NOT archival optical images (H$\alpha$, green; [N~{\sc ii}], red; [O~{\sc iii}], blue). The contours are color-coded as indicated in the legend (LSR velocities), with red, orange, cyan, and blue indicating a range from more highly redshifted to more highly blueshifted emission with respect to the systemic velocity of NGC 6445 ($V_{LSR} \sim +15$ km~s$^{-1}$).}
  \label{fig:contours}
\end{wrapfigure}

To illustrate how the ALMA line survey data are revealing the detailed molecular emission structures within these bipolar PNe, we present selected line (and continuum) emission maps for the PN NGC 6445  (Figs.~\ref{fig:mom0Mont}--\ref{fig:contours}). Fig.~\ref{fig:mom0Mont} provides a comparison between  optical (H$\alpha$) and ALMA ACA and 12-meter array images of the central region of NGC 6445. In the wider-field ACA maps, the molecular emission is seen to be confined to the central region of the nebula, precisely outlining the PN's bright, central H$\alpha$ ring. The ACA maps hence establish that the molecules previously (spectroscopically) detected in NGC 6445 with single-dish mm-wave facilities \citep{Schmidt2022} are in fact confined to an equatorial torus within the PN. The higher-resolution 12-m array maps reveal the detailed, clumpy morphology of this equatorial molecular torus. The positions and structures of individual molecular clumps within the torus seen in the 12-m maps closely correspond to features within the central ionized ring traced by H$\alpha$.
The detection of CN and CS molecules within the clumpy central region of NGC 6445 furthermore suggests the PN is rich in carbon, sulfur, and nitrogen.

Fig.~\ref{fig:VelocityChan} displays velocity channel maps extracted from the ACA $^{12}$CO (2--1) data cube. The same data are presented in the form of contour plots of emission integrated over specific velocity ranges, overlaid on the Nordic Optical Telescope (NOT) H$\alpha$ image, in Fig.~\ref{fig:contours}. The knotty structure of NGC 6445's molecular torus is even more apparent in these velocity-resolved representations.
Overall, as one moves from redshifted ($V_{\rm LSR}$ $\sim$ $+$25 to $+$45 km s$^{-1}$) to blueshifted ($V_{\rm LSR}$ $\sim$ $+$5 to $-$15 km~s$^{-1}$) emission, the knot system is observed to shift position around the torus from the northwest to the southeast of the central star, with knots at intermediate velocities (i.e., near the PN's systemic velocity, $V_{\rm LSR} \sim +15$ km s$^{-1}$) found at the limbs of the torus closest to its major axis. 



The radial velocity pattern in the ACA $^{12}$CO data thus demonstrates that NGC 6445's complex system of molecular knots, taken together, trace an expanding equatorial torus within the nebula. Furthermore, given the structures of the molecular knots as revealed in the 12-m array maps, it appears this torus is in the midst of being torn apart by the central star's high-energy radiation and/or fast winds. It seems reasonable to speculate that, as the nebula continues to evolve and the torus expands, it will further disintegrate, perhaps eventually resembling the highly fragmented systems of molecular clumps observed in the oldest Table 1 PNe, NGC 2818 and NGC 2899.

\section{Preliminary Conclusions and Future Prospects}

The same basic spatio-kinematic signature of an expanding molecular torus observed within NGC 6445 (Figs.~\ref{fig:VelocityChan}, ~\ref{fig:contours}) is also seen, to greater or lesser extents, in the ALMA data we have obtained for each of the Table~1 nebulae. Our initial ALMA results hence provide new evidence supporting the notion that the molecular gas reservoirs within bipolar PNe --- which presumably can be ascribed to high rates of mass loss near the end of the asymptotic giant branch evolutionary stages of the progenitor stars --- are largely confined to the equatorial regions of the nebulae.  Furthermore, in the context of the dynamical ages estimated thus far from our ALMA mapping of the molecular emitting regions within the Table 1 objects, the implication is that these high-excitation bipolar PNe display a temporal sequence of increasingly fragmented equatorial molecular gas structures --- from the apparently nearly complete, massive tori in Hb 5 and NGC 6302, to the clumpy central molecular rings in NGC 2440 and NGC 6445, and (finally) to the ragged systems of molecular clumps within the pinched-waist regions of NGC 2818 and NGC 2899 (Fig.~\ref{fig:Band6PNe}). Evidently, as molecule-rich bipolar PN age and expand, the combination of intense UV and winds from their very hot and luminous stars disrupts and disperses their AGB-ejecta-derived molecular tori, over $\sim$$10^4$ yr timescales. Our ongoing analysis of the molecular gas structures, kinematics, and chemical compositions revealed by the ALMA data summarized here will be aimed, in part, at confirming and elaborating on this proposed sequence describing the physical and chemical evolution of the equatorial tori within bipolar PNe descended from massive progenitor stars.

{\bf Acknowledgements}: This research is supported by U.S. National Science Foundation (NSF) grant AST-2206033 to RIT as well as by the NSF through award SOSPADA-009 from National Radio Astronomy Observatories.

\end{document}